\documentclass[twocolumn,showpacs,preprintnumbers,amsmath,amssymb]{revtex4}
\usepackage{epsfig}

\def\bea{\begin{eqnarray}}
\def\eea{\end{eqnarray}}

\def\nn{\nonumber}

\def\f{\frac}

\begin{document}

\title{Random-phase reservoir and a quantum resistor: The Lloyd model}
\author{Dibyendu Roy$^1$, N. Kumar$^1$  }
\affiliation{$^1$Raman Research Institute, Bangalore  560080, India}
\date{\today} 

\begin{abstract}
  We introduce phase disorder in a 1D quantum resistor through the formal device of `fake channels' distributed uniformly over its length such that the out-coupled wave amplitude is re-injected back into the system, but with a phase which is random. The associated scattering problem is treated via invariant imbedding in the continuum limit, and the resulting transport equation is found to correspond exactly to the Lloyd model. The latter  has been a subject of much interest in recent years. This conversion of the random phase into the random Cauchy  potential is a notable feature of our work. It is further argued that our phase-randomizing reservoir, as distinct from the well known phase-breaking reservoirs, induces no decoherence, but essentially destroys all interference effects other than the coherent back scattering.   
\end{abstract}      

\maketitle
 
  The Lloyd model \cite{Lloyd, Thouless72, Ishii73, Shepelyansky86, Lifshitz88, Mudry98, Deych00, Deych01} is known to be one of the very widely used models of disorder for quantum-electronic systems. Indeed, very recently it has been the subject of detailed analysis for electronic transport in a quantum resistor providing deeper insights into the scaling ideas of localisation in a 1D system \cite{Deych00, Deych01}. In the Lloyd model for a tight-binding disordered system, the site-energies are taken to be distributed identically, independently, and randomly with a Cauchy probability distribution. The latter is a fat-tailed distribution with infinite variance. Its simple two-pole structure in the complex site-energy plane makes for an exact analytical treatment. In this work we show that the Cauchy site-energy disorder (i.e., the random site-diagonal potential) can be formally viewed as arising from a certain process of phase randomization. The latter is introduced through the formal device of `fake or side channels' distributed uniformly along the length of the 1D resistor wherein the out-coupled wave amplitude is re-injected back into the system, but with the proviso that its phase is shifted randomly over $2\pi$. Such a phase disorder or `dephasing'-- without causing decoherence -- has been invoked recently \cite{Pilgram06, Forster07} in the context of mesoscopic conductors for calculating the full-counting statistics. Our objective here, however is different, namely, to study how such a random-phase distribution leads to a `potential' disorder giving the Lloyd model. This phase-randomization is formally incorporated through an invariant imbedding treatment as known in the context of quantum transport in disordered conductors \cite{Kumar85, Pradhan94, Rammal87, Heinrichs86}, where the object of interest is an emergent quantity such as the reflection/transmission coefficient or equivalently the resistance/conductance. The evolution equation so derived for the emergent quantity (the reflection amplitude in our case) in  sample length is found to correspond exactly to the continuum limit of the Lloyd model. This emergence of the Lloyd model with a Cauchy-potential disorder arising from the phase randomization through our phase-reservoir is a striking result. It is further argued that our phase-randomizing reservoir, unlike the well known phase-breaking (decohering) reservoirs \cite{Buttiker85, Buttiker86, Mello00}, can not eliminate the coherent back scattering. The phase randomization considered here by us involves effectively  parallel addition of quantum resistors (as introduced originally in Ref. \cite{ Shapiro83}) via the scattering matrices providing out-coupling to the side channels. Of course strictly speaking, being `quenched' in  nature, it can cause no reservoir-induced decoherence. 

\begin{figure}[t]
\begin{center}
\includegraphics[width=2.75in]{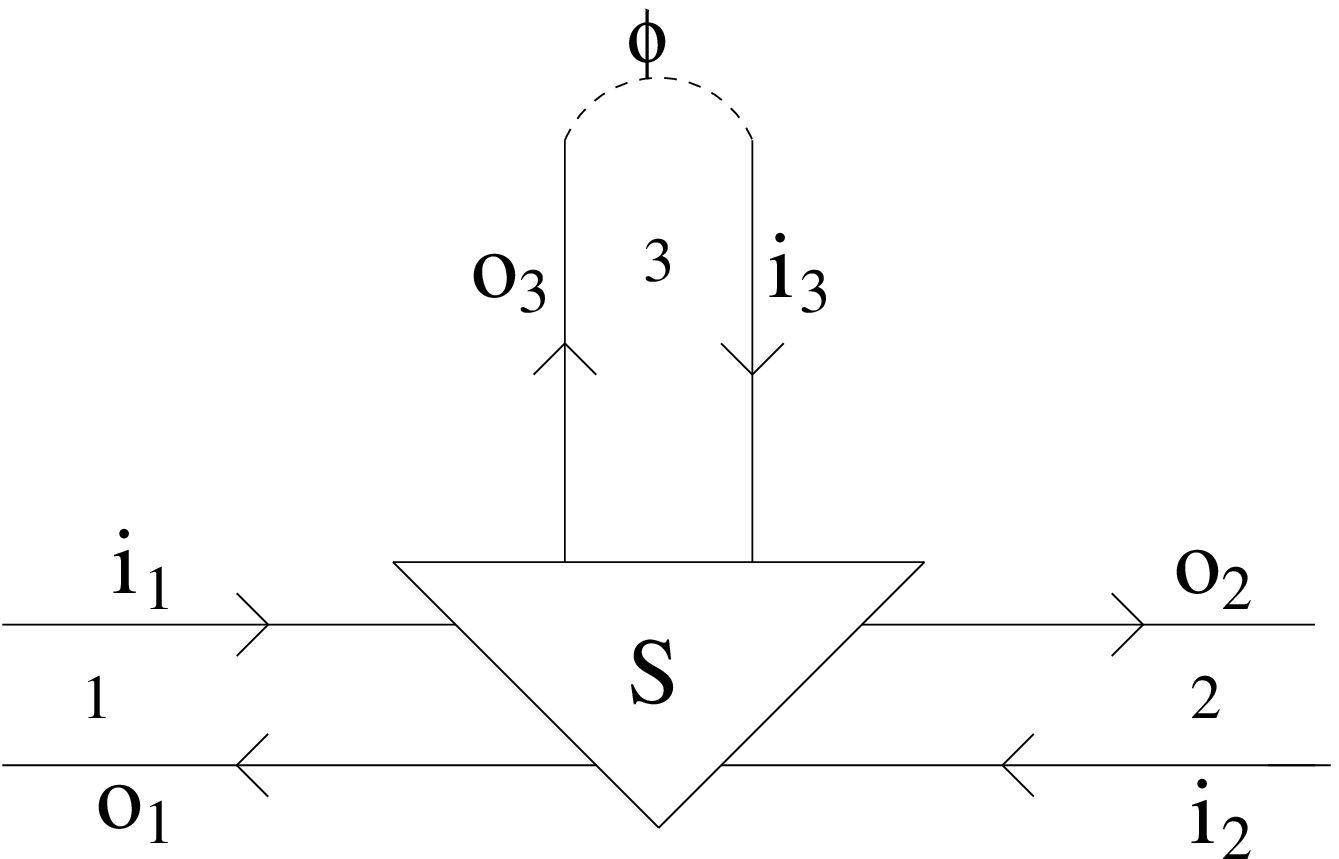}
\end{center}
\caption{ A schematic showing the random phase reservoir with the `fake channel' 3. Out-coupled amplitude is re-injected with random phase shift $\phi$}
\label{sPR}
\end{figure}

Let us first introduce our phase-randomizing reservoir with its `fake channels'. In its simplest form, it is modelled here by the three-port scatterer with an  energy-independent and symmetric $S$-matrix \cite{Buttiker85}  
\begin{eqnarray}
S = \left(
\begin{array}{lll}
 \f{1}{2}(\sqrt{1-2\epsilon}-1) &~~ \f{1}{2}(\sqrt{1-2\epsilon}+1)&~~~~~~ \sqrt{\epsilon}\\
\f{1}{2}(\sqrt{1-2\epsilon}+1) &~~ \f{1}{2}(\sqrt{1-2\epsilon}-1)&~~~~~~ \sqrt{\epsilon}\\
~~~~~~~~~ \sqrt{\epsilon} &~~~~~~~~~ \sqrt{\epsilon}&-\sqrt{1-2\epsilon}
\end{array}
\right)
\label{Smat}
\end{eqnarray}
connecting the outgoing amplitudes $(o_1, o_2, o_3)$ with the incoming amplitudes $(i_1, i_2, i_3)$ as shown in Fig.~\ref{sPR}.
 Here $\epsilon$ is the out-coupling to the transverse `fake channel' labelled 3 with $0 \leqslant \epsilon \leqslant \f{1}{2}$~. Channels 1 and 2 are the transport channels (leads) through which the device is to be inserted into the 1D quantum conductor. Our random-phase reservoir differs essentially from the well-known decoherence-inducing reservoirs \cite{Buttiker85, Buttiker86} in that the amplitude out-coupled into the `fake channel' is here re-injected (re-scattered) back into the system, but now with a phase shift $\phi$ which is assumed random over $2\pi$. 

\begin{figure}[t]
\begin{center}
\includegraphics[width=3.5in]{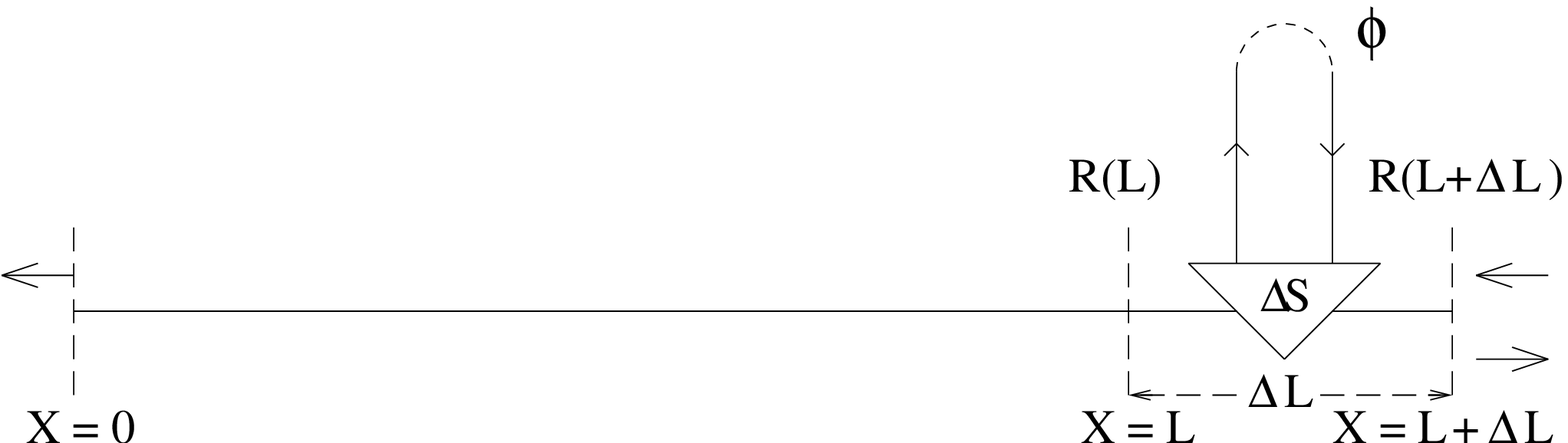}
\end{center}
\caption{ A schematic description of the `invariant imbedding' method for a 1D conductor with random-phase reservoirs distributed uniformly along the length.}
\label{Imbed}
\end{figure}

In order to introduce the random-phase reservoirs uniformly over the length of the 1D quantum resistor, we now use the method of invariant imbedding and solve the scattering problem for the emergent quantity (amplitude reflection coefficient in the present case). Following the general philosophy of invariant imbedding for a scattering problem, we now imbed the scattering sample of length $L$ in a super-sample of length $L+\Delta L$, and then study the change $\Delta S$ of the total $S$-matrix as $\Delta L$ tends to zero (Fig.~\ref{Imbed}). Here, $\Delta L$ contains the elementary random-phase reservoir  with the out-coupling $\epsilon$ of order $\Delta L$, i.e., $\epsilon/\Delta L \to~{\rm finite}~$ as $\Delta L \to 0$. Thus the parameter $\epsilon/\Delta L$  measures the strength per unit length with which the phase is randomized. The corresponding change $\Delta S$ in the $S$-matrix is then given by
\begin{eqnarray}
\Delta S = \left(
\begin{array}{lll}
~-\epsilon/2 &~~1-\epsilon/2 &~~~\sqrt{\epsilon}\\
1-\epsilon/2 &~~-\epsilon/2 &~~~ \sqrt{\epsilon}\\
~~~ \sqrt{\epsilon} &~~~~\sqrt{\epsilon}&-(1-\epsilon)
\end{array}
\right)
\label{delS}
\end{eqnarray}
In writing $\Delta S$ above we have made use of the fact that $\epsilon$ is small, of order $\Delta L$ in Eq.(\ref{Smat}). Next we calculate the incremental transmission ($\Delta T$) and the reflection ($\Delta R$) amplitudes in terms of the matrix elements (in obvious notation) $t_{13}=t_{23}=\sqrt{\epsilon},~t_{12}=1-\epsilon/2,~r_{33}=-(1-\epsilon)~{\rm and}~r_{11}=r_{22}=-\epsilon/2$ from the $\Delta$S above. Taking into account the multiple scatterings involving re-injection from the `fake channel', we obtain
\bea
\Delta T&=&t_{12}+t_{13}e^{i\phi}t_{32}+t_{13}e^{i\phi}r_{33}e^{i\phi}t_{32}+...\nn \\
&=&t_{12}+\f{t_{13}e^{i\phi}t_{32}}{1-r_{33}e^{i\phi}}\nn \\
&=&1-\f{\epsilon}{2}+\f{\epsilon e^{i\phi}}{1+(1-\epsilon)e^{i\phi}}~,\label{Tdl}\\
{\rm and}~~~~~~\Delta R&=&r_{11}+\f{t^2_{13}e^{i\phi}}{1-r_{33}e^{i\phi}}\nn \\
&=&\f{(e^{i\phi}-1)~\epsilon/2}{1+(1-\epsilon)e^{i\phi}}~.\label{Rdl}
\eea
Now, consider a plane wave incident on the right-hand side of the super-sample of length $L+\Delta L$. Summing over all processes of direct and multiple reflections and transmissions from the right-hand side of the sample of length $L$ and with the phase reservoir inserted in the interval [$L,~L+ \Delta L/2$], we have
\bea
R(L+\Delta L)= \Delta R + \f{{\Delta T}^2~e^{2ik\Delta L}~R(L)}{1-\Delta R~R(L)~e^{2ik\Delta L}}~,\label{IIb}
\eea
where $k$ is the wavevector magnitude for the incident electron wave. Expanding the right-hand side of Eq.(\ref{IIb}) using the values of $\Delta T~{\rm and}~\Delta R$ from Eqs.(\ref{Tdl},\ref{Rdl}), and keeping terms to order of $\Delta L$, we obtain 
\bea
\f{dR}{dl}=2iR(l)+\f{i}{2}\eta~\tan\f{\phi(l)}{2}(1+R(l))^2~,\label{Leq}
\eea
where we have introduced dimensionless length $l=kL$, and $\eta=\epsilon/k\Delta L$ as $\Delta L \to~0$, with the initial condition $R(l)= 0$ for $l = 0$~. Here the random phase $\phi(l)$ is distributed uniformly over 0 to 2$\pi$. Transforming $\eta\tan(\phi(l)/2) = V(l)$, we find the distribution $P_l(V)$ of $V(l)$
\bea
P_l(V) = \f{1}{\pi}\f{\eta}{V^2(l)+{\eta}^2},
\eea
which is the Cauchy probability distribution. Finally, with the above transformation from the random phase to the random potential (Cauchy), we obtain
\bea
\f{dR}{dl}=2iR(l)+\f{i}{2}V(l)(1+R(l))^2~.\label{CLeq}
\eea  
This invariant imbedding equation for evolution in $l$ has the form of a Langevin equation for the complex reflection amplitude $R$ with a Cauchy noise potential $V(l)$. It corresponds to the underlying quantum-mechanical Hamiltonian for a 1D disordered continuum with a potential $V(x),~0 \leqslant x \leqslant l$. The corresponding tight-binding Hamiltonian will have the site (Cauchy) potential $V(n)$ with $0 \leqslant n \leqslant N$, and $N=l/ka$ where $a$ is the lattice constant. Thus, the phase-randomization is mapped on to the Cauchy random potential $V(n)$ for a tight-binding Hamiltonian --- the Lloyd model. 
\begin{figure}[t]
\begin{center}
\includegraphics[width=2.75in]{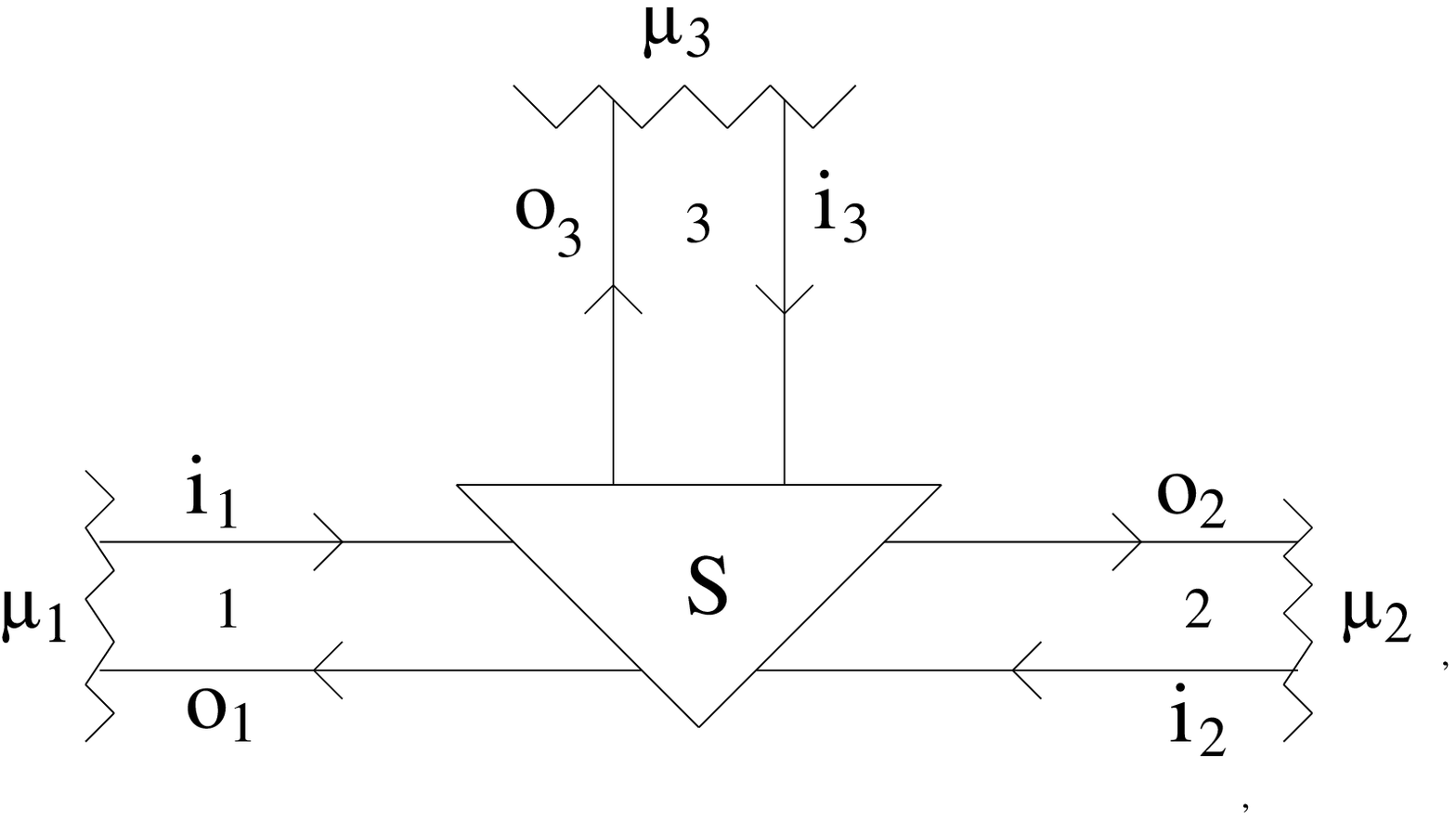}
\end{center}
\caption{ A schematic of the single-channel phase-breaking reservoir.}
\label{sBR}
\end{figure}

Having thus discussed the provenance of the Cauchy potential disorder (and, therefore, the Lloyd model) in terms of our random-phase reservoir, it will be in order now to compare the latter with the phase-breaking reservoirs giving the reservoir-induced decoherence, as due originally to B\"uttiker \cite{Buttiker85, Buttiker86}. For an isolated single-channel phase-breaking reservoir, the S-matrix is as given in Eq.~\ref{Smat}, and the corresponding schematic as in Fig.~\ref{sBR}. It shows explicitly the connections to the three terminals with  three chemical potentials: $\mu_1,~\mu_2$ for the longitudinal (or transport channels), and $\mu_3$ for the `potentiometric' (transverse) channel, the latter being determined from the condition of zero net current. This can be readily shown to give for the two-probe conductance ($G_{12}$) between terminals 1 and 2 
\bea
G_{12}=\f{e^2}{\pi\hbar}[(\f{1}{2}(\sqrt{1-2\epsilon}+1))^2+\f{\epsilon}{2}]. \label{SCon}
\eea 
In our corresponding random-phase reservoir with a single `fake channel', we have the same three-terminal S-matrix except for the re-injection at the `fake channel' 3 with a random phase $\phi$. For a given value of the phase $\phi$, the two-terminal conductance $G^{\phi}_{12}$ can be readily shown to be,
\bea
G^{\phi}_{12}=\f{e^2}{\pi \hbar}|t_{12} +\f{t_{13} e^{i\phi} t_{32}}{1-r_{33} e^{i\phi}}|^2~,  
\eea
with the coeffiecients $t_{12}=\f{1}{2}(\sqrt{1-2\epsilon}+1),~t_{13}=t_{32}=\sqrt{\epsilon}~{\rm and}~ r_{33}=-\sqrt{1-2\epsilon}$.
Averaging now $G^{\phi}_{12}$ over $\phi$, we find
\bea
\langle G^{\phi}_{12} \rangle_{\phi} \equiv \f{1}{2\pi}\int^{2\pi}_{0}G^{\phi}_{12} d\phi =G_{12}~,
\eea
 i.e., both the reservoirs give identical results for the two-probe conductance between the terminals 1 and 2. 

\begin{figure}[t]
\begin{center}
\includegraphics[width=2.75in]{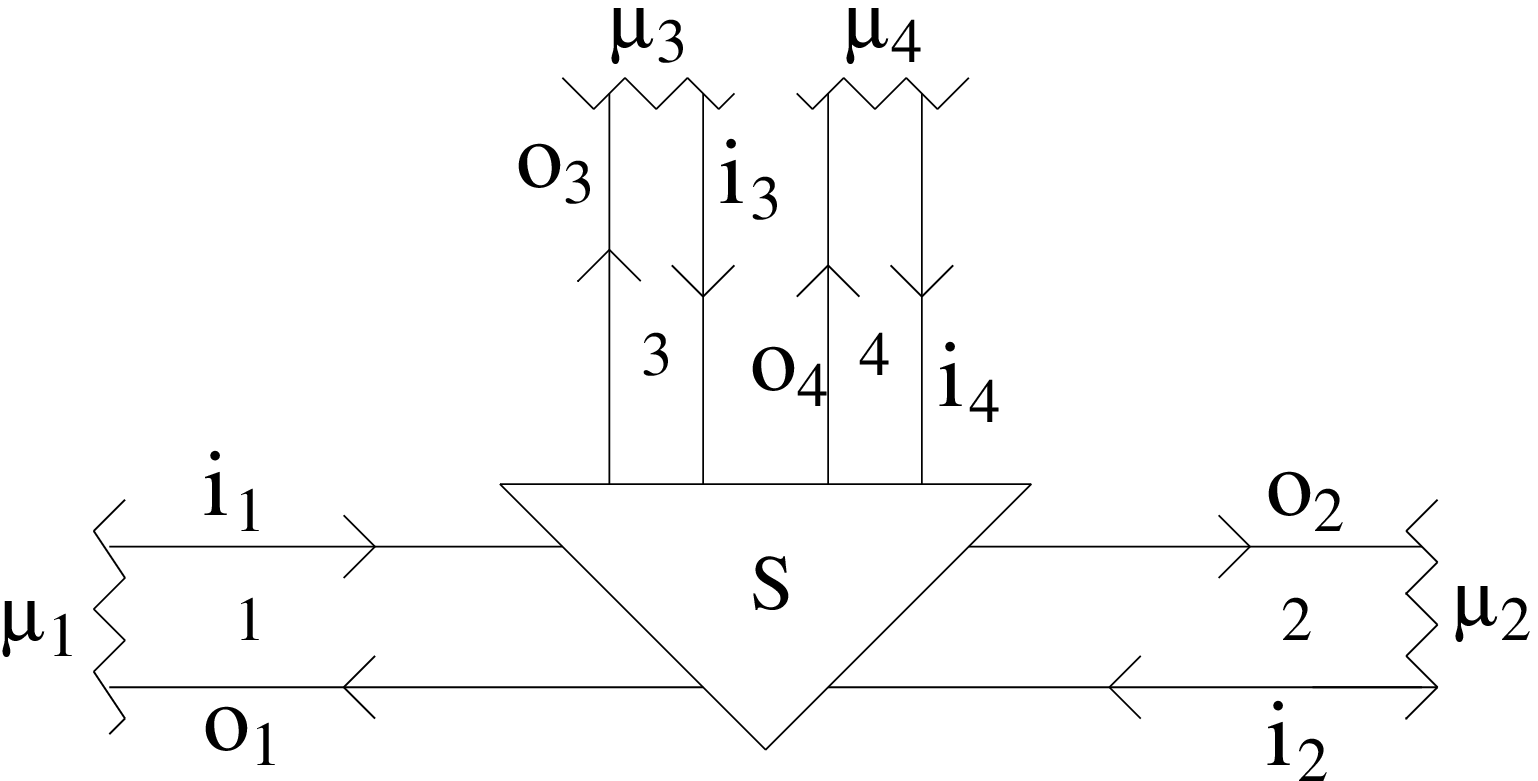}
\end{center}
\caption{ A schematic of  the phase-breaking reservoir with two un-coupled transverse channels 3 and 4.}
\label{dBR}
\end{figure}

\begin{figure}[t]
\begin{center}
\includegraphics[width=2.75in]{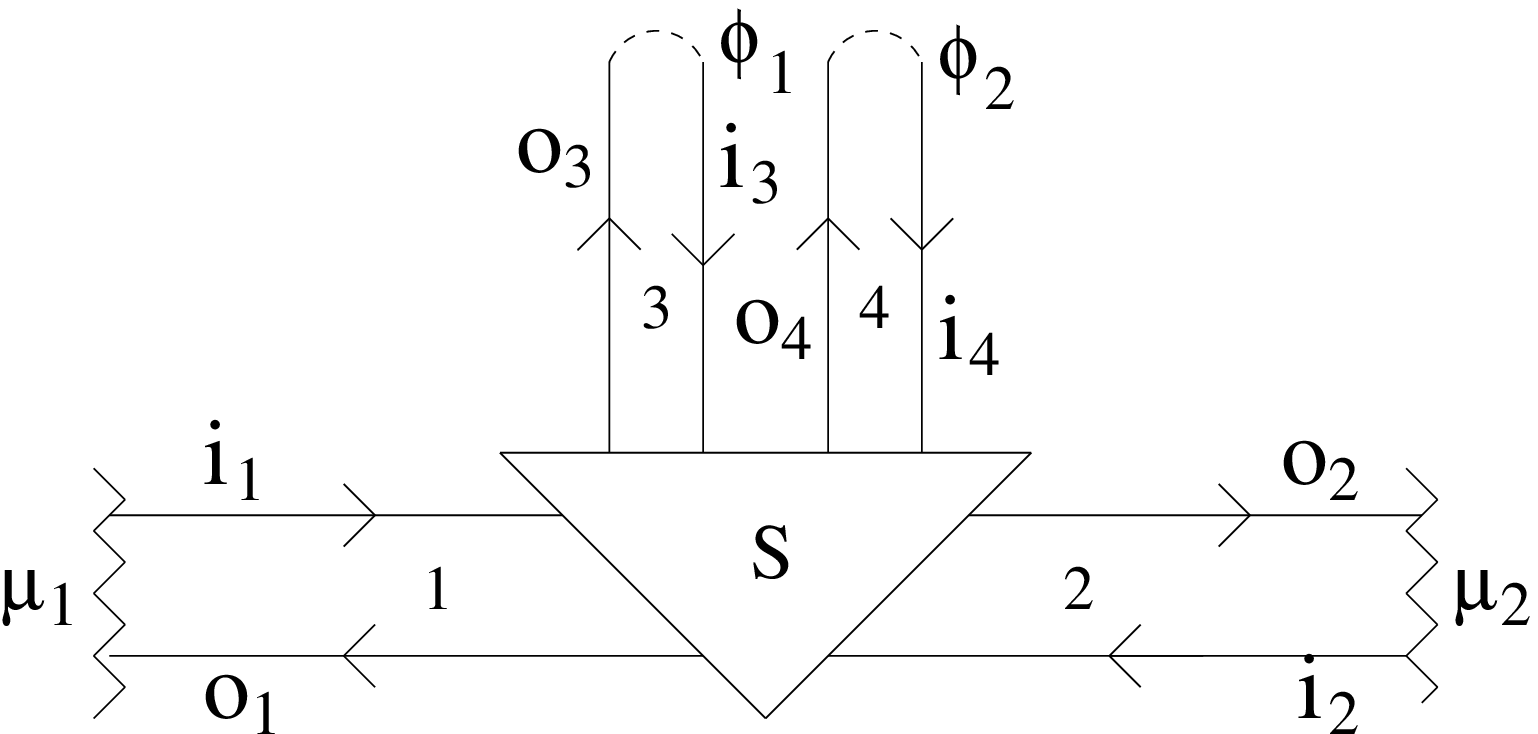}
\end{center}
\caption{ A schematic of the random-phase reservoir with two un-coupled `fake channels'.}\label{dPR}
\end{figure}

Now we turn to comparing the phase-breaking reservoirs with two transverse channels and our corresponding random-phase reservoir also with two `fake channels', as shown, in Figs.~(\ref{dBR},~\ref{dPR}). The corresponding 4-terminal  $S$-matrix is \cite{Buttiker86}
\begin{eqnarray}
S= \left(
\begin{array}{llll}
 ~~~~~0 &~~~ \sqrt{1-\epsilon} &~~~~~ \sqrt{\epsilon}&~~~~~~~0 \\
\sqrt{1-\epsilon} & ~~~~~~~0 &~~~~~~~0&~~~~~\sqrt{\epsilon} \\
~~~ \sqrt{\epsilon} &~~~~~~~0&~~~~~~~0&~-\sqrt{1-\epsilon}\\ 
~~~~~0 &~~~~~ \sqrt{\epsilon} &~- \sqrt{1-\epsilon}&~~~~~~~0
\end{array}
\right)
\end{eqnarray}
with $ 0 \leqslant \epsilon \leqslant 1$. Note the re-injections shown in dashes with random phases $\phi_1$ and $\phi_2$ at the `fake channels' 3 and 4 (Fig.~\ref{dPR}).
It is to be noted that in Fig.~(\ref{dBR} ) the `potentiometric' condition for zero net current is being imposed here for the two transverse channels 3 and 4 separately. With this, it can now be readily shown how that the two-probe conductances are again equal:
\bea
G_{12}=\langle G^{\phi_1,\phi_2}_{12} \rangle_{\phi_1,\phi_2}=\f{e^2}{\pi \hbar}\f{2(1-\epsilon)}{2-\epsilon}~.
\eea
Now, however, for the case of the two-channel phase-breaking reservoirs with the `potentiometric' condition of zero net current imposed summarily \cite{Buttiker86} on the two coupled transverse or side channels 3 and 4, the conductances turn out to be different. Some thought will convince that this is so because the phase-breaking reservoir and the random-phase reservoir differ essentially inasmuch as the former induces decoherence (can destroy all interference effects) while the latter can not eliminate the coherent back scattering (CBS). Indeed, for the case of coupled transverse channels, one an easily trace the CBS alternatives. We may say that our random-phase reservoir leads to a purification of interference effects to  coherent back scattering.

Now some comments and clarifying remarks on the use of the reservoirs in general, and the physical realization of the random-phase reservoir in particular as used here by us, seem to be in order. In the original Landauer-Buttiker scattering approach \cite{Landauer70, Buttiker88, Imry97} to quantum transport through a conductor, dissipation and associated decoherence are viewed as taking place in the reservoirs at the two ends of the sample. Physically, however, the dissipation takes place in the sample throughout its length. This latter feature has been modelled \cite{Buttiker85, Buttiker86, D'amato90, Roy07}, admittedly phenomenologically, through the formal device of reservoirs distributed along the sample length and connected to it through the appropriately chosen S-matrices whereby the out-coupled amplitude is absorbed and re-emitted into the sample where it adds incoherently to the coherent transport amplitude. This constitutes the now well-known reservoir-induced decoherence. Now, we can also have a random-phase reservoir where out-coupled amlitude is re-injected back into the conductor with a phase-shift distributed randomly over $2\pi$ as in the work presented here. We emphasize that this is a quenched phase-disorder that causes no decoherence or phase-breaking. The invariant imbedding in fact allows us to introduce both -- the decoherence \cite{Pradhan94, Pradhan06} as well as phase randomization -- over the conductor through a proper choice of $\Delta S~'s$ appearing in Eq.~(\ref{delS}), and calculate the emergent quantities like reflection/transmission coefficients. The random-phase reservoir is physically equivalent to the phase disorder as considered by some others \cite{Anderson80}. A literally physical realization of the random-phase reservoir would be through the chaotic cavities (with a long dwell time) terminating the side channels wherein the random phase-shifts result from the deterministic quantum chaos \cite{Pilgram06, Forster07}. The idea underlying the use of these formal devices (reservoirs) is that the strength of the out-couplings can be used to effectively parametrize some of the physical effects of interest. 

In conclusion, we have demonstrated analytically a conversion of random phases into random potentials that correspond exactly to the Lloyd model. To this end, we have introduced a formal device of random-phase reservoir with `fake channels'. Despite the apparent similarity to the well-known phase-breaking reservoirs, the two types are essentially different. Thus, while the phase-breaking reservoirs with absorption and re-emission of electrons cause the well known reservoir-induced decoherence (that can suppress all interference effects), our random-phase reservoirs having `fake channels' subtending re-injection with random phases, can not eliminate the coherent back scattering.


\end{document}